\documentclass[runningheads]{llncs}

\usepackage[T1]{fontenc}
\usepackage{graphicx}

\begin{document}

\title{A Large Dataset of Spontaneous Speech with the Accent Spoken in São Paulo for Automatic Speech Recognition Evaluation}
\titlerunning{A Large Dataset of Spontaneous Speech ...}

\author{Rodrigo Lima\inst{1}\orcidID{0009-0009-4344-1109} \and
Sidney E.\ Leal\inst{1}\orcidID{0000-0002-8817-2063} \and
Arnaldo Candido Junior\inst{2}\orcidID{0000-0002-5647-0891} \and
Sandra M.\ Aluísio\inst{1}\orcidID{0000-0001-5108-2630}
} 

\authorrunning{Rodrigo Lima  et al.}

\institute{University of São Paulo, São Carlos, SP 13566-590, Brazil \\
\email{guico21@usp.br, sidleal@gmail.com, and sandra@icmc.usp.br} \\
\and
 Universidade Estadual Paulista, São José do Rio Preto, SP, 15054-000, Brazil \\
\email{arnaldo.candido@unesp.br} \\
}

\maketitle

\begin{abstract}
We present a freely available spontaneous speech corpus for the Brazilian Portuguese language and report preliminary automatic speech recognition (ASR) results, using both the Wav2Vec2-XLSR-53 and Distil-Whisper models fine-tuned and trained on our corpus.  The NURC-SP Audio Corpus comprises 401 different speakers (204 females, 197 males) with a total of 239.30 hours of transcribed audio recordings. To the best of our knowledge, this is the first large Paulistano accented spontaneous speech corpus dedicated to the ASR task in Portuguese. 
We first present the design and development procedures of the NURC-SP Audio Corpus, and then describe four ASR experiments in detail. The experiments demonstrated promising results for the applicability of the corpus for ASR. Specifically, we fine-tuned two versions of Wav2Vec2-XLSR-53 model, trained a  Distil-Whisper model using our dataset with labels determined by Whisper Large-V3 model, and fine-tuned this Distil-Whisper model with our corpus. Our best results were the Distil-Whisper fine-tuned over NURC-SP Audio Corpus with a WER of 24.22\% followed by a fine-tuned versions of Wav2Vec2-XLSR-53 model with a WER of 33.73\%, that is almost 10\% point worse than Distil-Whisper's. To enable experiment reproducibility, we share the NURC-SP Audio Corpus dataset, pre-trained models, and training recipes in Hugging-Face and Github repositories.
\keywords{Automatic speech recognition evaluation \and Spontaneous speech \and Brazilian Portuguese \and Public speech corpora}
\end{abstract}

\section{Introduction}

Public or open datasets for training and evaluating automatic speech recognizers (ASR) in Brazilian Portuguese (BP) have increased in number and hours since mid-2020, when there were approximately 60 hours available, divided into four small datasets of read speech.
In 2024, more than 8 thousand hours are available for training ASR models, either from datasets automatically labelled or manually revised (see Table \ref{tab:asr} for a list of datasets). These BP resources allow the training of state-of-the-art ASR models, such as Wav2Vec2-XLSR-53 \cite{conneau21_interspeech} and Distil-Whisper \cite{gandhi2023distilwhisper}. Specifically, both models are suitable for low- or medium-resource languages, providing consistent results without the need for training sets having tens of thousand of hours. Wav2Vec2-XLSR-53 is pre-trained in a large dataset of unlabeled speech encompassing 53 languages, including BP, through auto-supervised learning. The pre-training process allows for efficient fine-tuning on available BP datasets, generating competitive ASR results. Distil-Whisper is a distilled version of Whisper \cite{radford23_whisper}. The latter, trained on 680,000 hours of multilingual data in a multitask approach, has shown superior performance across multiple datasets and domains when compared to speech recognizers using the Wav2Vec2 model. A disadvantage of the various large Whisper models is their requirement for a large dataset for increased performance and the inability to use them in computationally limited environments. Distil-Whisper, on the other hand, is a small variant based on knowledge distillation focused on efficiency which is 6 times faster, 49\% smaller, and performs within 1\% word error rate (WER) on out-of-distribution evaluation sets. To the best of our knowledge, Distil-Whisper has never been evaluated for BP datasets.

Regarding the data used for training speech recognizers, \cite{Gabler_2023} characterizes them in two dimensions: (i) \textbf{production style}, which describes a continuum of spontaneity, coming from one side, which is planned speech, to the other, which would be unplanned speech; (ii) \textbf{mode}, which characterizes the elicitation process of the training pairs, in which on the one hand the text is used as a stimulus for speech (leading to a read speech) and on the other hand it is the speech that is transcribed into text. Thus, read speech is planned speech that comes from a recited text and, at the other extreme, spontaneous/conversational speech is characterized as speech transcribed in an unplanned style. As a compromise, we have an speech prepared to be spoken later (e.g. the Ted Talks\footnote{https://www.ted.com/talks}, in which exponents from all over the world present a talk in 18 minutes or less, resulting in the need to use a planned text to transmit the message in the short time).

Spontaneous speech has phenomena that make its recognition more complex than that of read or prepared speech. As a result, datasets with read/prepared speech were first and widely disseminated for training speech recognizers or speech synthesizers, mainly for the English language (see \cite{zen2019libritts}, \cite{librispeech_2015}), since the words in the audios correspond directly to the words of the transcript, as they were read.
However, more naturally occurring conversational and spontaneous speech contains many prosodic phenomena that are not present in read speech, such as higher degrees of segmental reduction\footnote{For example, \cite{Bohn_2017} comments on vowel elision between words that, in the São Paulo dialect, affects the final posttonic vowels /a/, /o/ and /u/. For instance, in the example (a)    \textit{me’ren[da es]co’lar} (school lunch) --> \textit{me’ren[des]co’lar}, the vowel /a/ is deleted and a new syllable is created ([des]).} and more complex forms of variation (fillers, self-corrections and repetitions) affected by speech rate.

Thus, the various types of spontaneous speech that are common in everyday life, such as classes, conversations between two or more people, and interviews, are important materials for the datasets used to train ASRs, as they bring natural intonation to questions, statements, and expressions of emotions such as surprise, admiration, indignation, anger, astonishment, fright, exaltation, enthusiasm, among others \cite{szekely19b_interspeech,beckman1997typology}. Furthermore, they bring linguistic phenomena such as filled pauses, generally written as ``eh'', ``ah'', ``ahh'', ``mm'', ``uhn'' and editing disfluencies (repetitions of words or parts of words, revisions of what is intended to be said, with speech restarts) \cite{Liu2006}.
Consequently, speech recognizers are expected to perform worse with spontaneous speech than when applied to read speech. \cite{radford23_whisper} shows the application of ASR Whisper to 14 datasets of read and spontaneous speech in the English language. The highest WER (Word Error Rate) values occur for the spontaneous speech datasets, for example CHiME6\footnote{https://openslr.org/150/} has a WER of 25.5 and AMI SDM1\footnote{https://groups .inf.ed.ac.uk/ami/corpus/overview.shtml} has a WER of 36.4 while Common Voice\footnote{https://commonvoice.mozilla.org/en/datasets} has a WER of 9.0 and Tedlium\footnote{https://www.openslr.org/51/} has a WER of 4.0, using the large-v2 model. For Brazilian Portuguese, \cite{gris2023evaluating} reports a WER of 14.50\% for the Whisper large-v2 model applied to a Portuguese dataset of approximately 17 hours of spontaneous speech in interviews about life histories.

This paper introduces a new corpus of spontaneous BP speech, particularly of paulistano accent (São Paulo city), suitable for training and evaluating ASR systems. The NURC-SP Audio Corpus is part of São Paulo division of the NURC project  (\textit{Norma Urbana Linguística Culta}), a project designed to document and study Portuguese spoken language by people with a high degree of formal education in five Brazilian capitals (see details of this corpus in Section 3). 
It contains 239.30 hours of audios sampled at 16 kHz and their respective transcriptions, totalling  170k segmented audios. The audios were automatically transcribed for the first time and manually revised aiming at the ASR task. Therefore, this new corpus adds 239 hours to the amount available for BP (see Table 1), totaling 8,598 hours for training and evaluating speech recognition systems.

In order to compare the quality of our corpus, four ASR models are made available in this work: (i)  a fine-tuned version of Wav2Vec2.0 XLSR-53 with the train and validation subsets of NURC-SP  Audio Corpus, (ii) the same as the first, but  using as start point the model that \cite{candido_jr_et_al_2023} trained for CORAA-ASR v1.1\footnote{https://github.com/nilc-nlp/CORAA}, (iii)  a distilled version of Whisper Large-v3 model \cite{radford23_whisper}, a model that has support for the Portuguese language, trained using our dataset with labels determined by Whisper Large-v3, and (iv)  a fine-tuned version of the third model  with the train and validation subsets of NURC-SP  Audio Corpus.

The corpus and trained models are publicly available in our Github repository\footnote{github.com/nilc-nlp/nurc-sp-audio-corpus} under the CC BY-NC-ND 4.0 license.  The main contributions made in this paper are summarised as follows:

\begin{enumerate}
    \item A large BP corpus of human validated audio-transcription pairs containing 239.30 hours of spontaneous speech.
    \item The first corpus, according to our knowledge, tackling a large amount of accented paulistano speech for ASR in BP (CORAA ASR brings 31.14 hours of speech from São Paulo capital)
    \item Four ASR models, publicly available, based on the presented corpus.
\end{enumerate}

\section{Related Work on Datasets of Brazilian Portuguese for ASR}

Table 1 presents large corpora for building ASR systems focused on the Portuguese language. Some resources are multilingual, however Table 1 specifically details the statistics for the Portuguese language. Among the resources presented, there is a slightly greater preponderance of the Brazilian variant in the existing resources, although the European Portuguese variant is also included in some of them.

\begin{table}[ht]
\caption{Statistics of the main datasets available for ASR in Portuguese. Some datasets are multilingual, however, the numbers shown are for the Portuguese language.}
\centering
\label{tab:asr}
\resizebox{\textwidth}{!}{
\begin{tabular}{l|l|l|l|l|l}
\hline
\textbf{Corpora (Launching Date)} & \textbf{Speaking Style} & \textbf{Hours}  & \begin{tabular}[c]{@{}l@{}}\textbf{Number of}\\ \textbf{Audios}\end{tabular} & \begin{tabular}[c]{@{}l@{}}\textbf{Number of}\\ \textbf{Speakers}\end{tabular} & \textbf{License}  \\ \hline

MultiLingual LibriSpeech (MLS) (2020) & read        & 130.1   & -  & 54       & CC BY        \\
Multilingual TEDx (2021)       & prepared     & 164   & 93,000  & -        & CC BY-NC-ND 4.0   \\
Spotify Podcast Dataset (2022) & spontaneous  & 7,600 & 123,000 & -        & proprietary dataset   \\ 
CORAA ASR 1.1   (2022)          & spontaneous & 290   & 402,466 & 1,689    & CC BY-NC-ND 4.0    \\
Common Voice 17.0 (2024)        & read        & 175   & -       & 3,453    & CC-0          \\
\hline \hline
NURC-SP Audio Corpus (2024)     & spontaneous & 239.30 & 177,224 & 401     & CC BY-NC-ND 4.0   \\ \hline
\end{tabular}
}
\end{table}

One of the best-known projects dealing with read/prepared speech in English is Librivox\footnote{https://librivox.org/pages/about-librivox/}, which distributes public domain books in audio format. These audios were used in several projects to create resources for processing speech in English, such as LibriSpeech ASR Corpus\footnote{https://www.openslr.org/12} and LibriTTS\footnote{https://www .openslr .org/60/}, both hosted in the Open Speech and Language Resources repository.

The MultiLingual LibriSpeech (MLS) corpus \cite{pratap20_interspeech} is a large multilingual corpus suitable for speech research, derived from read audiobooks from LibriVox and consists of 8 languages, including about 32K hours of English and a total of 4.5K hours for other languages. For Portuguese, there are 131 hours and 54 speakers. Specifically for the recognition task, it can be combined with other resources, as it has relatively few speakers (audiobook speakers).  These resources consist of cleaner audio, generally in studio quality. Because of this, models built solely on this type of audio are only suitable for speech recognition in low-noise scenarios. To overcome this feature, one solution would be to inject noise into the audio or combine it with audio from other projects at different quality levels.

The MultiLingual TeDx Corpus \cite{salesky21_interspeech} was proposed to enable research in the areas of automatic speech recognition and speech-to-text translation. For Portuguese, there are 164 hours available in 93k audios. The corpus is made up of talks on a wide range of subjects, being managed within the scope of the TEDx project, linked to the TED group (Technology, Entertainment and Design). In the case of Portuguese, there are also translations of the transcriptions into English and Spanish. In addition, audios in Spanish and French also have translations into Portuguese.

The Spotify corpus \cite{clifton-etal-2020-100000} was first released in the English language. In 2022, the company launched a new version\footnote{https://arxiv.org/abs/2209.11871} incorporating Portuguese \cite{cem_mil_portugues_2023}, offering several audios for the Portuguese language coming mainly from podcasts available on the platform. In total, 76k hours of audio were made available from 123k episodes of shows on the platform. Transcripts were automatically generated and are subject to transcription errors. It has a free license for academic use, but researchers interested in accessing the audios must submit a request for access on the organizers' website.

CORAA ASR\cite{candido_jr_et_al_2023} is a corpus for automatic speech recognition that contains specially spontaneous speech. CORAA ASR is the combination of five independent projects dealing with speech in the interior of São Paulo (with 35.96 hours --- ALIP Project), Minas Gerais (with 9.64 hours ----- C-ORAL Brasil Project), Recife (141.31 hours --- NURC-Recife Project) and São Paulo capital (31.14 hours --- SP2020 Project), in addition to of the prepared speech of TeDx Talks in Brazilian and Portuguese Portuguese (72.74 hours), totaling 290 hours and 402k audios.

The Common Voice corpus \cite{ardila-etal-2020-common} is an open-use project created by the Mozilla Foundation. The project is a response to the lack of resources for several languages, including Portuguese. In the project, users can simultaneously contribute to the growth of the base and access other people's audio.  To collaborate with the project, users can donate audio in their own voices and review donations from other users. The project has tools for collection, validation and internationalization (adaptation to different languages). The permissive use license of this project allows the exploration of the corpus including for commercial purposes. In version 17, the subcorpus for the Portuguese language has 211 hours of audio and transcriptions, of which 175 were validated.

The NURC-SP Audio Corpus, the focus of this paper, is described in detail in Section 3.

\section{The NURC-SP Corpus}

NURC-SP was the São Paulo division of the NURC (\textit{Norma Urbana Linguística Culta}) project. NURC-SP collected more than 300 hours of São Paulo speakers throughout the 1970s. The NURC-SP corpus is made up of 375 audio recordings of three genre types: 

\begin{enumerate}
\item (DID) Documenter and Participant - Dialogue carried out between a documenter and a participant directly; 
\item (D2) Participant and Participant - Dialogue between two participants mediated by a documenter; and 
\item (EF) Participant - Lectures, seminars, classes, speeches in general, given by a participant in a formal context.
\end{enumerate}

\noindent which are divided into three subcorpora: the Minimum Corpus (21 audio recordings),  the Corpus of Non-Aligned Audios and Transcriptions (26 audio recordings), and the Audio Corpus.
    
NURC-SP had its original analog audios digitized by the Alexandre Eulalio Cultural Documentation Center (CEDAE/UNICAMP) and in December 2020 it was made available to the Tarsila Project\footnote{\url{https://sites.google.com/view/tarsila-c4ai/home}} as a resource to:  (i) build training data sets for spontaneous speech recognition systems, and (ii) facilitate future linguistic studies.  The three subcorpora that integrate the NURC-SP Digital repository were made available in the Portal NURC-SP Digital \cite{rodrigues-etal-2024-portal} for researchers from fields of Linguistics as well for the general public, due to easy access and filtering tools to use the material.

Here, in this paper, we focus on the NURC-SP Audio Corpus. It was originally composed of 328 audio recordings without transcriptions, which have been automatically transcribed by WhisperX \cite{bain2022whisperx}. The revision of automated WhisperX transcription segments were performed from June 2023 to December 2023 by 14 native speakers of BP. The revision process was based on an annotation guideline designed to: (i) help making the revision uniform and (ii) remove segments with high amount of noise and overlapping voices. The guideline contains 11 rules, dealing with (i) orality marks, (ii) how to transcribe filled pauses, (iii) repetitive hesitations, (iv) numbers, (v) individual letters, (vi) acronyms, (vii) foreign terms, (viii) punctuation and capitalization, (iv) emotion sounds (ex. laughter) which were annotated in parentheses, (x) misunderstanding of words or passages and (xi) how to deal with automatic segmentation failures.

\subsection{Pre-processing and filtering of NURC-SP Audio Corpus for evaluating ASR Models}

Audio quality was a very important characteristic when preparing the dataset for evaluating ASR models. In this section, we present the processing steps of the NURC-SP Audio Corpus to generate the version used to evaluate the four ASR models in this paper:

\begin{enumerate}
\item Removal of those audios (and all of its segments) with poor/fair quality
(with high number of problems with voice distortion, hissing, background noise and interruptions);
\item Annotation of the segments using ``high/low'' quality labels for all remaining audios, based on what was reported by human annotators during the automatic transcription revision;
\item Removal of part of the low quality segments, as detailed further in this section, in order to generate the final dataset for training ASR models; and
\item Generation of both statistics of NURC-SP Audio Corpus and the train, validation and test sets.
\end{enumerate}

Of the 328 audios, four of them had poor quality, 92 had fair quality, 52 had good quality and 180 had excellent quality.
Therefore, of these 328 audios, five of them were not used for training the ASR models in this paper: SP\_EF\_395, SP\_DID\_283, SP\_DID\_194, SP\_D2\_397, SP\_D2\_337, as they had problems with voice distortion, hissing, background noise and interruptions, as they were responsible for generating transcriptions with a high rate of errors.
   
The transcriptions were processed to remove the labels used in the automatic transcription revision: (i) ``\#\#\#'' was used to indicate segments with high background noise, very low audio, overlapping voices and music; (ii) paralinguistic sounds such as laughter or cough were annotated in parentheses --- (laughter), (cough); (iii)  misunderstanding of words or passages were also marked with parentheses ``( )''; and (iv) words truncated at the end or the beginning of the audio due to automatic segmentation failure were partially transcribed with the help of  ``>''  and ``<'' (for example: if the word ``casa'' -- house) where truncated at the end of the segment, it was annotated as ``ca>'' and if it was at the beginning, it was annotated as ``<sa''.

Segments marked with ``\#\#\#'' are removed completely, as well as segments having only paralinguistic sounds (e.g., laughter) or containing only truncated words. Segments having truncated and non-truncated words (marked with ``>'' or ``<'') had the truncated word removed and were marked as low quality.
Pairs that have words or passages marked as misunderstood, using the ``( )'' tags, had the tags removed and the pairs were marked as ``low quality''. Audio-transcription pairs with paralinguistic sounds tags only had the tags removed and were marked as ``high quality''.
The dataset available in this work has 323 audio-transcription pairs, which were divided into 177,224 segments transcribed by WhisperX, and revised by native BP speakers (see more details in Table \ref{tab:statistics}).

\subsection{Statistics of NURC-SP Audio Corpus}

Table \ref{tab:statistics} presents statistics of NURC-SP Audio Corpus Dataset. Overall, almost 240 hours of speech were generated in the final export, distributed in around 177 thousand segments with an average duration of approximately 5 seconds. In total, more than 2 million tokens were transcribed, resulting in approximately 12 tokens per segment. It should be noted that the sum of audios with females and males is higher than the total number of audios, because a given audio can have simultaneously both types of voices.

\begin{table}
\centering
\caption{NURC-SP Audio Corpus Statistics}
\label{tab:statistics}
\begin{tabular}{l|c|c|c|c}
\hline
& \hspace{0.2cm} \textbf{Training} \hspace{0.1cm} & \hspace{0.2cm} \textbf{Validation}  & \hspace{0.2cm} \textbf{Test} \hspace{0.2cm} & \hspace{0.2cm} \textbf{Total} \hspace{0.2cm} \\
\hline
Number of Audios (unique)                        & 303            & 6            & 14           &  323  \\
Female voices (number)*                       & 191            & 4            & 9           & 204   \\
Male voices (number)*                       & 186            & 3            & 8           & 197   \\
Male/Female Ratio              &  0.97           & 0.75            & 0.89           & 0.97  \\

Duration (hours)                       & 224.47            & 4.60            & 10.23           & 239.30   \\
Quantity of Audios (segmented)                       & 166,971            & 3,142            & 7,111           & 177,224   \\
Segment Duration (avg seconds)                       & 4.83            & 5.27            & 5.18           &  4.86   \\
Segment Duration (max seconds)                       & 29.87            & 29.70            & 28.98           &  29.87   \\
Avg Tokens & 11.81 & 12.67 & 12.32 & 11.85 \\
Avg Types & 10.61 & 11.73 & 12.29 & 10.69 \\
Total Tokens & 1,971,993 & 39,715 & 87,598 & 2,099,306 \\
Total Types & 84,767 & 8,005 & 12,218 & 88,004 \\
Type/Token Ratio & 0.043 & 0.202 & 0.139 & 0.042 \\
\hline
\end{tabular}
*There are audios with two speakers, in several combinations: Male and Male, Male and Female, Female and Female.
\end{table}

\begin{figure}[!htb]
\centering
\includegraphics[scale=1.2]{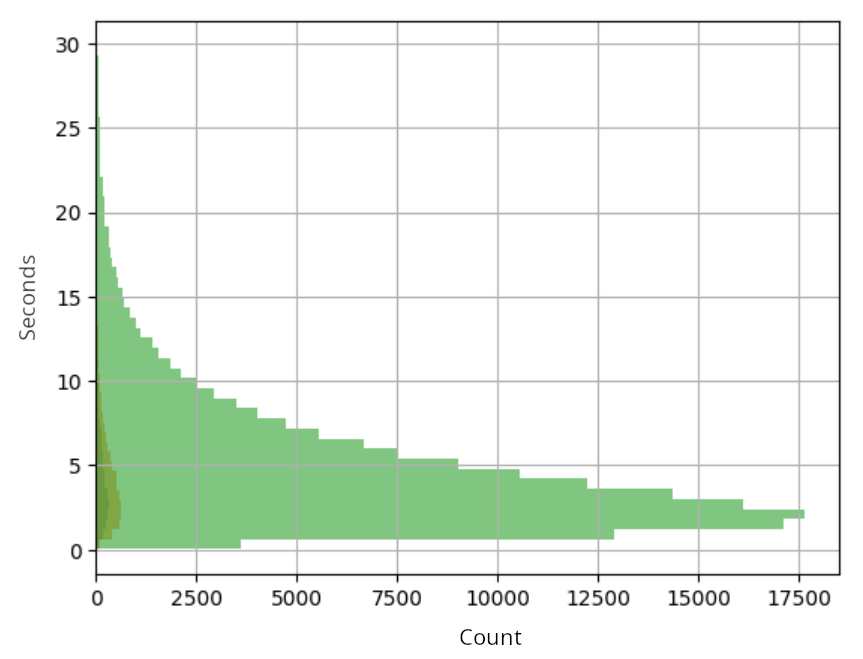}
\caption{Audio Segments Duration}
\label{fig:statistic_duration}
\end{figure}

Figure \ref{fig:statistic_duration} shows the duration of audio segments. It can be noted that audios smaller than 5 are predominant, with a peak in approximately 3 seconds, while the interval from 5 to 20 seconds is also well represented. There are also audios longer than 20s, but their occurrence is less common.

\begin{figure}[!htb]
\centering
\includegraphics[scale=0.46]{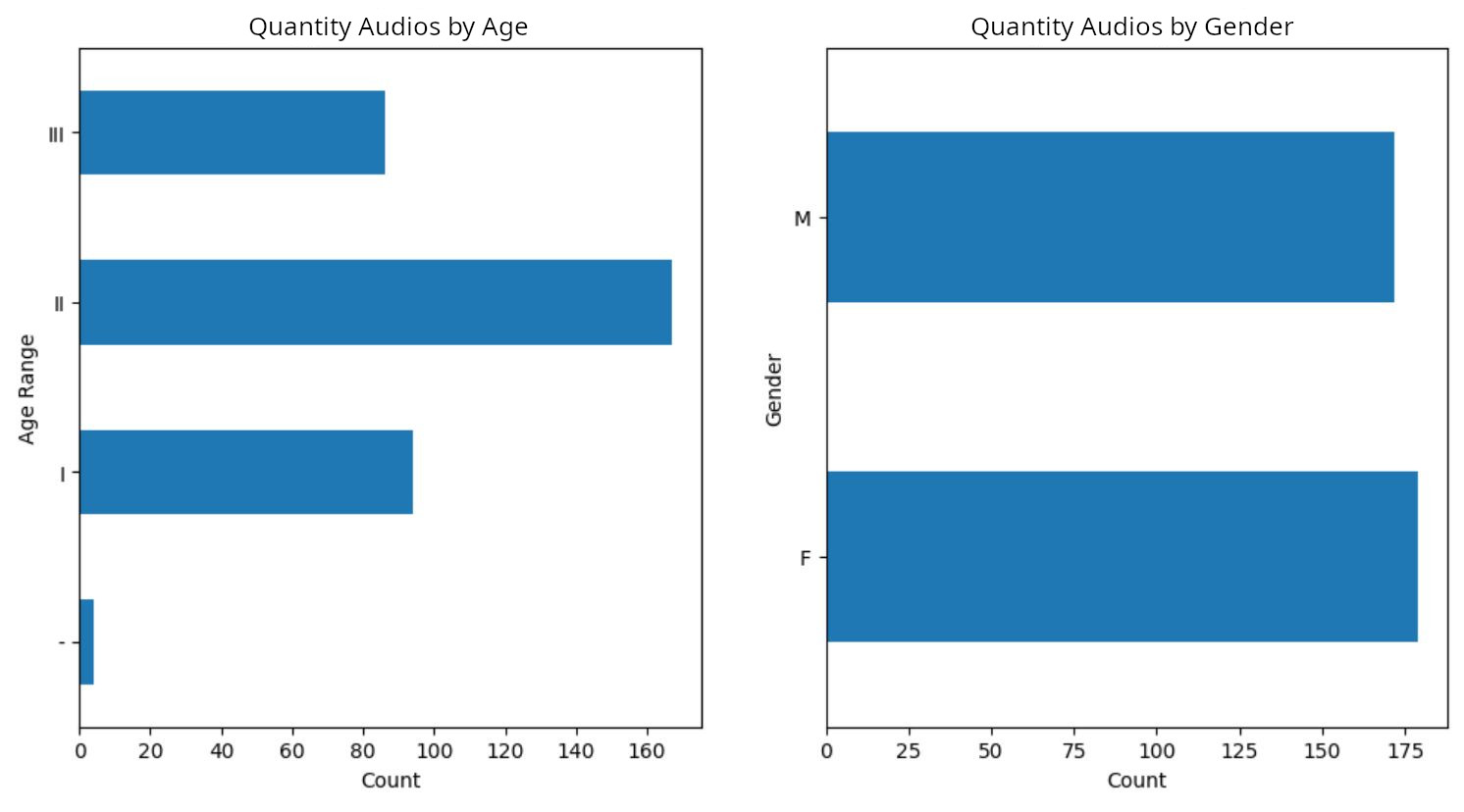}
\caption{(a) Audios by Speaker's Age Range and (b) Gender (M/F). Speakers are distributed into three age groups (I = 25–35, II = 36–55, and III = 56 onwards).}
\label{fig:statistic_age-range-gender}
\end{figure}

Figure \ref{fig:statistic_age-range-gender} shows the number of audios by age and gender. The number of audios of females and males are similar, resulting reasonable balancing regarding gender, which can also be seem in male/female ratio in Table \ref{tab:statistics}. Regarding age, there are more speakers in age group II (from 36 to 55 years old) than the others groups (I=25--35 and III=36-55). There also a small number of audios with speakers of unknown age.

\begin{figure}[!htb]
\centering
\includegraphics[scale=0.48]{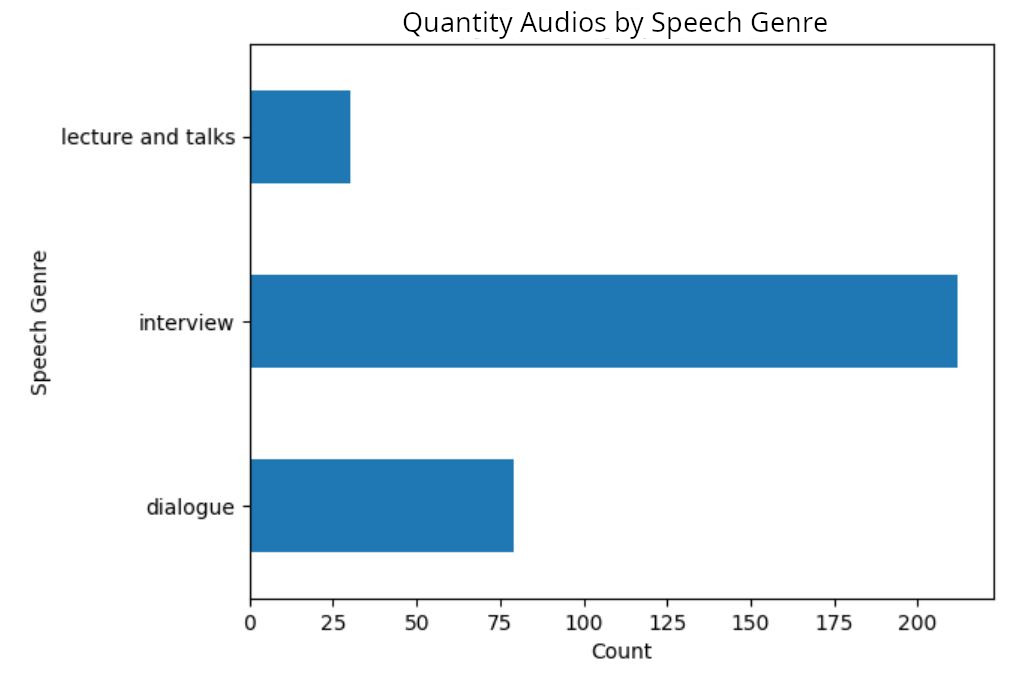}
\caption{Audios by Speech Genre. Lectures and Talks receive the acronym EF, Interviews are called DID and Dialogues  between two participants are named D2.}
\label{fig:statistic_speech-range}
\end{figure}

Figure \ref{fig:statistic_speech-range}
shows the number of audios by speech gender (Lectures and Talks --- (EF), Interviews (DID) and Dialogs (D2), respectively). Interviews and dialogues are much more common than lectures, totalling 90\% of the audios. Thus, NURC-SP is a corpus predominantly made of spontaneous speech.

\section{Experiments on NURC-SP Audio Corpus}

We performed experiments over NURC-SP Audio Corpus in order to assess the dataset quality and limitations.
We based our experiments in two main architectures: Wav2Vec2 \cite{conneau21_interspeech} and Distil-Whisper \cite{gandhi2023distilwhisper}. Wav2Vec2 has the advantage of presenting a good performance in low and middle resource languages. Wa2Vec2 is an auto-supervised pretraining in different languages, which speeds up the training process for ASR, allowing robustness against noises that are common in spontaneous speech. Distil-Whisper \cite{gandhi2023distilwhisper} is a distilled version of Whisper, the state-of-the-art in ASR in several languages. We opted for the distilled version because original Whisper is costly, being originally trained over thousands hours of labelled speech, while Distil-Whisper can perform efficiently with less training data. Distil-Whisper is smaller and faster, while preserving many of original Whisper's strengths.

Wav2vec2 architecture operates directly on the wave signal, without the need of intermediary representations such as spectrograms. Furthermore, there are two main layer types: convolutional layers process the raw wave signal; Transformer \cite{vaswani2017attention} layers process the result of the previous step. The model can be trained in an auto-supervised manner. The signal is also quantized in the after passing through convolutional layers and masked language modelling is used in the Transformer layers. A rich loss takes into account quantized representations and the Transformer layer predictions in order to produce diversified quantized representations that represent small windows in the original audio. The learned representations can then be fine-tuned to represent, for example, phonemes in the process of ASR. In this case, it is used a classical seq2seq loss for ASR generation, namely the CTC loss \cite{graves2006connectionist}.

Whisper proposes an off-the-shelf approach to its architecture, consisting on a traditional transformer, with encoding and decoding blocks, and focusing more on the training data than on architectural adjustments. The input must be converted to log-mel spectrogram format. Distil-Whisper is the distilled version of Whisper. Model distillation is a technique that allows for a smaller model (the student) to be trained on both data labels as well as predictions from a bigger model (the professor) in a given prediction task. The main idea behind distillation is that a model's predictions have richer information about the data distribution than original labels. Therefore, a bigger model can capture data distribution and transmit it to the student. Both Whisper and Distil-Whisper are available in different versions. Comparing Whisper Large-v3 with Distil-Whisper Large-v3 (used in this work), it can be observed an improvement of 5.8 times in inference time while using 51\% fewer parameters in the distilled version.

Table \ref{tab:datasets_training} presents statistics of training, validation and test partitions of each NURC-SP Audio Corpus subset.

\begin{table}[!ht]
\centering
\caption{Statistics of training, validation and test partitions of each NURC-SP Audio Corpus subset: the total duration in hours and number of audio recordings (left) and  number of speakers in each gender (right).}
\begin{tabular}{c|c|c|c|c|c|c}
\hline
\multicolumn{1}{c|}{\textbf{Subset}} & \multicolumn{3}{c|}{\textbf{Hours/Number}} & \multicolumn{3}{c}{\textbf{Speakers (M|F)}}           \\ \hline
\multicolumn{1}{c|}{}    & Training       & Validation    & Test            & \hspace{0.1cm} Training \hspace{0.1cm}            & \hspace{0.2cm} Validation \hspace{0.2cm}           & Test           \\ \hline
\multicolumn{1}{l|}{D2}        & 75.93/75      & 1.45/1      & 3.12/3        &  66|82           &  1|1          &  3|3               \\ \hline
\multicolumn{1}{l|}{DID}       & 132.39/205      & 1.46/2      & 3.83/6        &  103|103         &  1|1          &  3|3               \\ \hline
\multicolumn{1}{l|}{EF}        & 16.15/23      & 1.69/3      & 3.87/5       &   17|6           &  1|2          &  2|3              \\ \hline
\multicolumn{1}{l|}{\textbf{Total}}     & 224.47/303      & 4.60/6      & 10.23/14      &  186|191         &  3|4        &  8|9             \\  \hline
\end{tabular}
\label{tab:datasets_training}
\end{table}

\subsection{Baseline Models Development}
To assess the dataset quality, we trained four new ASR models and compared the results with previous related works. Two of them are fine-tuned versions of Wav2Vec2 and the third and fourth models are approaches with Distil-Whisper. All four models are described below.

\textbf{Wav2Vec2-NURC-SP-1.}
This model is a fine-tuned version of Wav2Vec 2.0 XLSR-53 \cite{baevski_et_al_2020} \cite{conneau_et_al_2020}, pre-trained over 53 languages (Portuguese included). The pre-trained model was fine-tuned with our train and validation subsets of (NURC-SP Audio Corpus) NURC-SP-AC  in one GPU Nvidia DGX A100 80GB for 16 epochs, with early stop of 10. The other settings were the same from \cite{candido_jr_et_al_2023}, training code is available at github\footnote{\url{https://github.com/Edresson/Wav2Vec-Wrapper}}.

\textbf{Wav2Vec2-NURC-SP-2.}
The second model is almost the same as the first one, but using as start point the model that \cite{candido_jr_et_al_2023} trained for CORAA-V1 and made publicly available at HuggingFace\footnote{\url{https://huggingface.co/Edresson/wav2vec2-large-xlsr-coraa-portuguese}}. It also was trained for 16 epochs and finished with a better WER that the first model.

\textbf{Distil-Whisper-NURC-SP.}
The third model is a distilled version of Whisper Large-v3 \cite{radford23_whisper}, a model that has support for the Portuguese language, trained using our dataset with labels determined by Whisper Large-v3, with the reason being that more knowledge can be passed from the teacher model to the student model this way \cite{gandhi2023distilwhisper}. The model was trained with our train and validation subsets of NURC-SP-AC in one GPU Nvidia DGX A100 80GB for 48 epochs, following to the steps given by the Distil-Whisper github\footnote{\url{https://github.com/huggingface/distil-whisper/tree/main/training}}.

\textbf{Distil-Whisper-NURC-SP-Fine-Tuned.}
This model is a fine-tuning of the third one with our train and validation subsets of NURC-SP-AC, following the steps recommended by the Distil-Whisper developers\footnote{\url{https://huggingface.co/blog/fine-tune-whisper}}. It was also trained with 48 epochs in one GPU Nvidia DGX A100 80GB and achieved a better WER and CER (Character Error Rate)\footnote{Here, we also focus our analysis on the metric CER, 
because for smaller audios, with just a few words, this metric tends to be more reliable.} than the third model.

\subsection{Normalization}

We performed the following steps to normalize the train, validation and test subsets into a standardized form in order to penalize only when a word error is caused by failure in transcribing the word, and not by formatting, punctuation or spontaneous speech differences.
The revised transcriptions of NURC-SP Audio Corpus are of different text genders (EF, D2 and DID) and use upper and lower case letters and punctuation, as well as filled pause markers such as \textit{eh}, \textit{hum}, \textit{ãh}, etc. To simplify the training and calculation of the CER and WER metrics, the following normalization was performed:
\begin{enumerate}
\item The texts were transformed into lowercase;
\item All punctuation marks generated by Whisper were removed (ellipsis, exclamation mark, fullstop, question mark, and comma);
\item The filled pauses were standardized to: \textit{eh, uh, ah}, as follows: \textit{eh = eh, éh, ehm, ehn}; \textit{uh = uh, hm, uhm, hmm, mm, mhm}; \textit{ah = ah, huh, ãh, ã};
\item Any successive blank spaces have been replaced with one space.
\end{enumerate}

\subsection{Experiments Results}

For comparison purposes, we kept the tests on the datasets used in previous works and added the new dataset made available by this work. The \cite{gris_et_all_2022} and \cite{candido_jr_et_al_2023} models were rerun on all datasets, as were the four new trained models. WER and CER metrics were evaluated for each run. The new dataset proved to be quite challenging for the ASR task, as can be seen in Table \ref{tab:results_models}.

\begin{table}[!ht]
\centering
\caption{NURC-SP Audio Corpus Baseline Models results compared with previous works. The best values for CER and WER appear in bold.}
\begin{tabular}{l|c|c|c|c|c|c|c|c}
\hline
\textbf{Datasets}                     & \multicolumn{2}{|c|}{  \begin{tabular}[c]{@{}c@{}}\textbf{Common}\\ \textbf{Voice}\end{tabular} } & \multicolumn{2}{c|}{  \textbf{CORAA v1 } } & \multicolumn{2}{c|}{  \begin{tabular}[c]{@{}c@{}}\textbf{NURC-SP}\\ \textbf{Audio Corpus}\end{tabular} } & \multicolumn{2}{c}{\textbf{Mean}} \\
\hline
Metrics                       & CER            & WER            & CER           & WER           & CER          & WER          & CER         & WER        \\
\hline
Gris et al. (2022) \cite{gris_et_all_2022}           & \textbf{4.50}           & \textbf{16.32}          & 22.32         & 43.70          & 26.52         & 47.74         & 17.78        & 35.92       \\
Candido Jr et al. (2023) \cite{candido_jr_et_al_2023} & 6.99           & 24.44          & \textbf{11.02}         & \textbf{24.18}         & 22.87         & 40.29         & 13.63        & 29.64       \\
Wav2Vec2-NURC-SP-1                         & 10.41           & 35.74           & 24.24          & 49.13          & 23.69         & 43.44         & 19.45        & 42.77     \\
Wav2Vec2-NURC-SP-2                         & 8.07           & 26.74           & 14.59          & 31.19          & 19.30         & 33.73         & 13.99        & 30.55     \\
Distil-Whisper-NURC-SP                         & 7.14           & 18.66           & 23.53          & 36.17          & 25.03         & 36.25         & 18.57        & 30.36     \\
\begin{tabular}[c]{@{}l@{}}Distil-Whisper-\\ NURC-SP-Fine-Tuned \end{tabular} &      5.70      &     17.76       &      14.89     &     26.91      & \textbf{15.77}         & \textbf{24.22}         &    \textbf{12.12}     &   \textbf{22.96}   \\
\hline
\end{tabular}
\label{tab:results_models}
\end{table}

Below, we show four examples of result instances from our best model (Distil-Whisper  
NURC-SP-Fine-Tuned) --- the first is the Original Normalized (ON) and the second is the Model's Prediction Normalized (MPN).
In the first two, there are problems with named entities as the  ``Martinelli'' building was transcribed as ``martini'' and the name ``Paulo Emilio Salles Gomes'' had a wrong transcription as ``paulo e milho fales gomes''. The model  used more frequent common names than the proper names  ``Martinelli'' and ``Emilio Salles''.  Rare words related to certain domains (e.g. food) or terminologies of research areas (e.g. linguistics) also bring some difficulties to the model (see third and fourth examples):

\begin{enumerate}
    \item (ON) \textit{o martinelli ficou célebre em todo o exterior do estado no interior do estado de são paulo e mesmo pelo brasil afora como um arranha-céu notável para a época}; (MPN) \textit{o martini ficou célebre em todo o exterior do estado do interior de estado de são paulo e mesmo pelo brasil afora como arranha-céu notável para a época};
    \item (ON) \textit{você me falou em cinema eu lembrei de paulo emilio salles  gomes que foi meu colega na faculdade e é um entendidíssimo de cinema né}; (MPN) \textit{você me falou em cinema eu me lembrei de paulo e milho fales gomes que foi minha colega  na faculdade e é um entendidíssimo de cinema né};
    \item (ON) \textit{cuscuz paulista bobó de camarão essas coisas assim}; (MPN) \textit{cuscos paulista babota de camarão essas coisas};
    \item (ON) \textit{de um lado objeto direto do outro adjunto}; (MPN) \textit{de um lado é o chefe do e o outro é de junho}.
\end{enumerate}

\section{Discussion and Conclusions}
In this paper, we present a freely available spontaneous speech corpus for the Brazilian Portuguese language, totaling 239.30 hours,
and report preliminary ASR results, using both the Wav2Vec2-XLSR-53 and Distil-Whisper models fine-tuned and trained on our corpus. To the best of our knowledge, Distil-Whisper has never been evaluated for BP datasets. 

For comparison purposes, we also bring the tests on the datasets used in previous works.
The Wav2vec 2.0 based model from \cite{gris_et_all_2022} remains the best for the Common Voice dataset with a WER of 16.32\%, even in the most recent version (CV version 17.0), as it is trained in prepared/read speech. In the CORAA v1 dataset, \cite{candido_jr_et_al_2023}'s models, which is also trained an Wav2Vec 2.0 model, but mostly based on spontaneous speech, continues presenting the best values for this dataset with 24.18\% WER, but our Wav2Vec2-NURC-SP-2 model was in second position (since it is also trained for spontaneous speech). 
For the new dataset, focus of this paper, the best results were the Distil-Whisper fine-tuned over NURC-SP Audio Corpus with a WER of 24.22\% followed by a fine-tuned versions of Wav2Vec2-XLSR-53 model with a WER of 33.73\%, that is almost 10\% point worse than Distil-Whisper's. These results indicates that Distil-Whisper is promising for low and medium resource languages and should be evaluated with more BP datasets in the future.

\begin{credits}
\subsubsection{\ackname} This work was carried out at the Center for Artificial Intelligence (C4AI-USP), with support by the São Paulo Research Foundation (FAPESP grant \#2019/07665-4) and by the IBM Corporation.  This project was also supported by the Ministry of Science, Technology and Innovation, with resources of Law No. 8.248, of October 23, 1991, within the scope of PPI-SOFTEX, coordinated by Softex and published Residence in TIC 13, DOU 01245.010222/2022-44.

\end{credits}

\bibliographystyle{splncs04}   
\bibliography{references}   
\end{document}